\documentclass[twocolumn,aps,pra,longbibliography]{revtex4-1}

\usepackage{amsmath,amssymb,bm,color,graphicx}

\newcommand{\imag}{\text{Im}}

\begin{document}

\title{Equivalence between quantum backflow and classically forbidden probability flow in a diffraction-in-time problem}

\author{Arseni Goussev}

\affiliation{Department of Mathematics, Physics and Electrical Engineering, Northumbria University, Newcastle Upon Tyne NE1 8ST, United Kingdom}

\date{\today}

\begin{abstract}
	Quantum backflow is an interference effect in which a matter-wave packet comprised of only plane waves with non-negative momenta exhibits negative probability flux. Here we show that this effect is mathematically equivalent to the appearance of classically-forbidden probability flux when a matter-wave packet, initially confined to a semi-infinite line, expands in free space.
\end{abstract}

\maketitle


\section{Introduction}

Quantum mechanics admits states of matter, known as ``backflowing'' states, that possess the following counter-intuitive property: even though a momentum measurement performed on a backflowing state is guaranteed to return a non-negative value, the outcome of a probability flux measurement may be negative. The existence of such a backflow for a linear superposition of two (non-normalizable) plane waves was first identified in Refs.~\cite{All69time-c} and \cite{Kij74time}. In 1994, Bracken and Melloy reported the first in-depth theoretical study of the effect for normalizable wave packets~\cite{BM94Probability}. One of their surprising discoveries was the fact that the maximal amount of the probability that can possibly flow in the ``wrong'' direction is a constant $c_\text{bm}$ that is independent of Planck's constant or of any system parameter and has a numerical value of approximately 4\%. (As of today, the exact value of the backflow constant remains unknown; the most accurate numerical estimate states $c_\text{bm} \simeq 0.0384517$ \cite{PGKW06new}.) Consequently, Bracken and Melloy proposed that $c_\text{bm}$ is a ``new dimensionless quantum number''. The desire to better understand the nature of the backflow effect, and to explore its manifestations in various physical processes has been at the center of numerous investigations \cite{MB98velocity, EFV05Quantum, PGKW06new, Ber10Quantum, Str12Large, YHHW12Analytical, PTMM13Detecting, HGL+13Quantum, BCL17Quantum, EZB--Observation}.

The standard formulation of the backflow problem proceeds as follows. One considers the time-evolution, in free space~\cite{BM94Probability} or under the action of a constant force~\cite{MB98velocity}, of a quantum state confined (at all times) to non-negative momenta, $p \ge 0$, but unconstrained in position space, $-\infty < x < +\infty$. Then, finding a space-time point $(x,t)$ where the probability flux $J(x,t)$ is negative constitutes an effect that is impossible from the viewpoint of classical mechanics. In this paper, we show that there is a one-to-one correspondence between the backflow problem outlined above and another problem in non-relativistic quantum mechanics hereinafter referred to as the problem of ``quantum reentry''. In the quantum reentry problem, one is interested in the free-space propagation of a non-relativistic wave packet which is initially, at $t = 0$, localized to the non-positive semi-infinite position axis, $x \le 0$, but is unconstrained in momentum space, $-\infty < p < +\infty$. In the course of time, the wave packet expands into the positive position region, $x > 0$, and one addresses its probability flux $\mathcal{J}(x,t)$ at $x = \ell \ge 0$ and $t > 0$. It is clear that the corresponding classical-mechanical probability flux at $\ell \ge 0$ must remain non-negative for all $t > 0$, which represents the fact that a free classical particle, originated from the region $x \le 0$, can only arrive at point $\ell$ with a positive velocity. In quantum mechanics however there is nothing that prevents the particle from reentering the $x < \ell$ region from the right, rendering $\mathcal{J}(\ell,t)$ negative at some $t > 0$, a phenomenon well known from the studies of arrival time in quantum theory (see, e.g., Ref.~\cite{ML00Arrival} for a review). What seems to be unknown, and what we establish in the present paper, is that the equations governing the classically-forbidden probability transfer in the quantum reentry scenario are the same as in the quantum backflow case. In particular, the maximal value of the total probability that can pass the point $x = \ell$ from right to left in the quantum reentry problem appears to be equal to the maximal backflow probability for a positive-momentum wave packet moving under the action of a constant force; in the limit $\ell \rightarrow 0$, the reentry probability approaches $c_\text{bm}$.

The problem of the time-evolution of a free quantum particle initially confined to a semi-infinite line has a long history. In 1952, Moshinsky considered the free-space propagation of an initially ``chopped'' monochromatic beam of non-relativistic particles represented by the wave function $\theta(-x) e^{i k x}$, where $k > 0$ and $\theta(\cdot)$ denotes the Heaviside step function. He discovered that the probability flux $\mathcal{J}(\ell,t)$ at a spatial point $\ell > 0$ oscillates in time $t$, and that these oscillations are mathematically analogous to the intensity fringes observed in the Fresnel limit of diffraction of light at the edge of a straight, semi-infinite screen~\cite{Mos52Diffraction}. This analogy prompted Moshinsky to term the flux-oscillation phenomenon as ``diffraction in time'' (DIT). Since then, DIT has been the topic of intense theoretical and experimental research (see Ref.~\cite{CGM09Quantum} for a review and Refs.~\cite{Mou10Diffraction, TMB+11Explanation, GH12Delay, Gou13Diffraction, BD15Three, GV16basis, GCAS17Single} for some more recent results). The question of finding the maximal reentry probability, addressed in this paper, is essentially Moshinsky's DIT problem in a variational context: for $\ell \ge 0$, one looks for a normalized initial wave function of the form $\theta(-x) \psi(x)$ that maximizes the probability transfer from the spatial region $x > \ell$ to the region $x < \ell$ during a time interval $\tau_1 < t < \tau_2$. As demonstrated below, this variational problem turns out to be mathematically equivalent to the variational problem arising in the study of quantum backflow against a constant force.

The paper is organized as follows. In Sec.~\ref{bf}, we review the effect of quantum backflow against a constant force; our presentation follows Ref.~\cite{MB98velocity}, with some minor modifications. In Sec.~\ref{dit}, we address the effect of quantum reentry and show that it is mathematically equivalent to that of quantum backflow. In Sec.~\ref{end}, we summarize our findings and discuss similarities and differences between the backflow and reentry effects.

\section{Quantum backflow against a constant force}
\label{bf}

The effect of quantum backflow against a constant force $m g$, where $m$ is the particle mass and $g$ is the particle acceleration, was originally studied in Ref.~\cite{MB98velocity}. Here we summarize key elements of the theoretical description of the effect and present them in a way that facilitates comparison between the quantum backflow and quantum reentry problems.

Consider the time-evolution of a non-negative-momentum wave packet $\Phi(x,t)$, initially given by
\begin{equation}
	\Phi(x,0) = \frac{1}{\sqrt{2 \pi \hbar}} \int_0^{\infty} dp \, \phi(p) e^{i p x / \hbar} \,, \label{bf-ini}
\end{equation}
that is governed by the Schr\"odinger equation
\begin{equation}
	i \hbar \frac{\partial \Phi}{\partial t} = -\frac{\hbar^2}{2 m} \frac{\partial^2 \Phi}{\partial x^2} - m g x \Phi \,, \qquad g \ge 0 \,. \label{bf-SE}
\end{equation}
The non-negativity of force $m g$ guarantees that the plane-wave decomposition of $\Phi(x,t)$ involves only plane waves with non-negative momenta for all $t \ge 0$. The wave packet is normalized throughout its time-evolution, i.e. $\int_{-\infty}^{\infty} dx \, |\Phi(x,t)|^2 = 1$; this condition is equivalent to the following constraint on the initial momentum distribution function, $\phi(p)$:
\begin{equation}
	\int_0^{\infty} dp \, |\phi(p)|^2 = 1 \,. \label{bf-norm}
\end{equation}
The probability flux associated with $\Phi(x,t)$ is given by
\begin{equation*}
	J(x,t) = \frac{\hbar}{m} \imag \left\{ \Phi^*(x,t) \frac{\partial \Phi(x,t)}{\partial x} \right\} \,,
\end{equation*}
where the asterisk denotes complex conjugation. Calculated at $x = 0$, the flux reads
\begin{align}
	J(0,t) = &\frac{1}{4 \pi \hbar m} \int_0^{\infty} dp \int_0^{\infty} dp' \, (p + p' + 2 m g t) \phi^*(p) \phi(p') \nonumber \\
	&\quad \times \exp \left( \frac{i t}{2 \hbar m} (p + p' + m g t) (p - p') \right) \,. \label{bf-flux}
\end{align}
The quantum backflow effect manifests itself in the following way. It appears to be possible to find momentum-space wave functions $\phi(p)$ such that $J(0,t) < 0$ for some range of $t > 0$. The overall probability transfer from the right half-line, $x > 0$, to the left half-line, $x < 0$, over a time interval $T_1 < t < T_2$ (with $T_1 \ge 0$) is given by
\begin{align}
	P
	&= -\int_{T_1}^{T_2} dt \, J(0, t) \nonumber \\
	&= \int_0^{\infty} dp \int_0^{\infty} dp' \, \phi^*(p) K(p,p') \phi(p') \,, \label{bf-P}
\end{align}
where
\begin{equation}
	K(p,p') = \frac{i}{2 \pi} \frac{\left. \exp \left\{ \frac{i t}{2 \hbar m} (p + p' + m g t) (p - p') \right\} \right|_{t = T_1}^{t = T_2}}{p - p'} \,. \label{bf-K}
\end{equation}
(In Ref.~\cite{MB98velocity}, $T_1$ is taken to be zero.) In terms of the dimensionless wave function
\begin{align*}
	f(z) = &\left( \frac{4 \hbar m}{T_2 - T_1} \right)^{1/4} \\
	&\times \exp \left( -i \frac{T_1 + T_2}{4 \hbar m} p^2 - i g \frac{T_1^2 + T_2^2}{4 \hbar} p \right) \phi(p)
\end{align*}
with
\begin{equation*}
	z = \frac{1}{2} \sqrt{ \frac{T_2 - T_1}{\hbar m} } \, p \,,
\end{equation*}
Equations~(\ref{bf-norm}) and (\ref{bf-P}) read, respectively,
\begin{equation}
	\int_0^{\infty} dz \, |f(z)|^2 = 1 \label{bf-norm2}
\end{equation}
and
\begin{align}
	P = -\frac{1}{\pi} &\int_0^{\infty} dz \int_0^{\infty} dz' \nonumber \\
	&\times f^*(z) \frac{\sin \left[ (z + z' + \alpha) (z - z') \right]}{z - z'} f(z') \,, \label{bf-P2}
\end{align}
where
\begin{equation}
	\alpha = g \sqrt{\frac{m}{\hbar} (T_2 - T_1)} \, \frac{T_1 + T_2}{2} \,. \label{alpha}
\end{equation}
The maximal backflow probability, obtained by maximizing $P$ over $f$ under the normalization constraint (\ref{bf-norm2}), equals the largest eigenvalue $\lambda = P_{\max}$ (or, more precisely, the supremum of the eigenvalue spectrum) in the following integral eigenvalue problem:
\begin{equation}
	-\frac{1}{\pi} \int_0^{\infty} dz' \, \frac{\sin \left[ (z + z' + \alpha) (z - z') \right]}{z - z'} f(z') = \lambda f(z) \,. \label{bf-eigenproblem}
\end{equation}
As of today, no (nontrivial) analytical solution to Eq.~(\ref{bf-eigenproblem}) is available. Numerical investigations of the eigenproblem (\ref{bf-eigenproblem}) suggest that the largest eigenvalue is given by $P_{\max} \simeq c_\text{bm} e^{-2 \alpha}$  \cite{MB98velocity}. In the limit $\alpha \rightarrow 0$, Eq.~(\ref{bf-eigenproblem}) reduces to the integral equation for the maximal backflow probability in free space~\cite{BM94Probability}, i.e. $P_{\max} = c_\text{bm}$ for $\alpha = 0$.

\section{Quantum reentry in free space}
\label{dit}

We now turn to the problem of quantum reentry -- the DIT-type problem concerned with the free motion of a quantum particle initially confined to a semi-infinite line, $x \le 0$. Here, the wave function $\Psi(x,t)$, starting from
\begin{equation}
	\Psi(x,0) = \theta(-x) \psi(x) \,, \label{dit-ini}
\end{equation}
evolves in accordance with the free-particle Schr\"odinger equation
\begin{equation}
	i \hbar \frac{\partial \Psi}{\partial t} = -\frac{\hbar^2}{2 m} \frac{\partial^2 \Psi}{\partial x^2} \,. \label{dit-SE}
\end{equation}
The wave function is normalized to unity, i.e.
\begin{equation}
	\int_{-\infty}^0 dx \, |\psi(x)|^2 = 1 \,. \label{dit-norm}
\end{equation}
At time $t > 0$, the wave function is given by
\begin{equation*}
	\Psi(x,t) = \int_{-\infty}^0 dx' \, U(x-x',t) \psi(x') \,,
\end{equation*}
where
\begin{equation*}
	U(\xi,t) = \sqrt{\frac{m}{2 \pi i \hbar t}} \exp \left( i \frac{m \xi^2}{2 \hbar t} \right)
\end{equation*}
is the free particle propagator.

The probability flux associated with $\Psi(x,t)$ is
\begin{equation*}
	\mathcal{J}(x,t) = \frac{\hbar}{m} \imag \left\{ \Psi^*(x,t) \frac{\partial \Psi(x,t)}{\partial x} \right\} \,.
\end{equation*}
A straightforward evaluation of the flux at a spatial point $x = \ell \ge 0$ and time $t > 0$ yields
\begin{align}
	\mathcal{J}(\ell,t) = &\frac{m}{4 \pi \hbar t^2} \int_{-\infty}^0 dx \int_{-\infty}^0 dx' \, (2 \ell - x - x') \psi^*(x) \psi(x') \nonumber \\
	&\quad \times \exp \left( \frac{i m}{2 \hbar t} (2 \ell - x - x') (x - x') \right) \,. \label{dit-flux}
\end{align}
At this stage, one can already see some similarity between the expression above for $\mathcal{J}(\ell,t)$ and that for $J(0,t)$, given by Eq.~(\ref{bf-flux}). This similarity suggests that it should be possible to find position-space wave functions $\psi(x)$ that would generate negative probability flux, $\mathcal{J}(\ell,t) < 0$, at some $t > 0$. Note that such a negative probability flux is impossible in the corresponding classical scenario in which a cloud of free non-interacting particles, initially localized in the semi-infinite region $x \le 0$, expands into the region $x > 0$. In other words, once a classical particle has left the region $x < \ell$ it can no longer reenter it.

The overall classically-forbidden reentry probability -- the probability transfer from the region $x > \ell$ to the region $x < \ell$ -- over a time interval $\tau_1 < t < \tau_2$ is given by
\begin{align}
	\mathcal{P}
	&= -\int_{\tau_1}^{\tau_2} dt \, \mathcal{J}(\ell,t) \nonumber \\
	&= \int_{-\infty}^0 dx \int_{-\infty}^0 dx' \, \psi^*(x) \mathcal{K}(x,x') \psi(x') \,, \label{dit-P}
\end{align}
where
\begin{equation}
	\mathcal{K}(x,x') = \frac{i}{2 \pi} \frac{\left. \exp \left\{ \frac{i m}{2 \hbar t} (2 \ell - x - x') (x - x') \right\} \right|_{t = \tau_2}^{t = \tau_1}}{x - x'} \,. \label{dit-K}
\end{equation}
(The time integral in Eq.~(\ref{dit-P}) is readily evaluated via the substitution $\nu = 1/t$.) Finally, the introduction of the dimensionless wave function
\begin{align*}
	f(z) = &\left( \frac{4 \hbar}{m} \right)^{1/4} \left( \frac{1}{\tau_1} - \frac{1}{\tau_2} \right)^{-1/4} \\
	&\times \exp \left[ i \frac{m}{4 \hbar} \left( \frac{1}{\tau_1} + \frac{1}{\tau_2} \right) (x - 2 \ell) x \right] \psi(x)
\end{align*}
with
\begin{equation*}
	z = -\frac{1}{2} \sqrt{\frac{m}{\hbar} \left( \frac{1}{\tau_1} - \frac{1}{\tau_2} \right)} \, x
\end{equation*}
transforms Eq.~(\ref{dit-norm}) into Eq.~(\ref{bf-norm2}), and Eq.~(\ref{dit-P}) into
\begin{align}
	\mathcal{P} = -\frac{1}{\pi} &\int_0^{\infty} dz \int_0^{\infty} dz' \nonumber \\
	&\times f^*(z) \frac{\sin \left[ (z + z' + \beta) (z - z') \right]}{z - z'} f(z') \,, \label{dit-P2}
\end{align}
where
\begin{equation}
	\beta = \ell \sqrt{\frac{m}{\hbar} \left( \frac{1}{\tau_1} - \frac{1}{\tau_2} \right)} \,. \label{beta}
\end{equation}
Clearly, Eqs.~(\ref{bf-P2}) and (\ref{dit-P2}) are identical. This shows that the problems of quantum backflow and quantum reentry are mathematically equivalent to each other. That is, for every $g$, $T_1$, $T_2$, and $\phi(p)$, yielding the backflow probability $P$, there are $\ell$, $\tau_1$, $\tau_2$, and $\psi(x)$, yielding the reentry probability $\mathcal{P} = P$; the converse is also true. It is worth nothing that the maximal reentry probability is given by the largest eigenvalue $\lambda = \mathcal{P}_{\max}$ (or, more precisely, by the supremum of the eigenvalue spectrum) in the integral eigenproblem (\ref{bf-eigenproblem}) with $\alpha$ replaced by $\beta$, i.e.
\begin{equation}
	-\frac{1}{\pi} \int_0^{\infty} dz' \, \frac{\sin \left[ (z + z' + \beta) (z - z') \right]}{z - z'} f(z') = \lambda f(z) \,. \label{dit-eigenproblem}
\end{equation}
It follows from the numerical investigation reported in Ref.~\cite{MB98velocity} that $\mathcal{P}_{\max} \simeq c_{\text{bm}} e^{-2 \beta}$. For $\beta = 0$, Eq.~(\ref{dit-eigenproblem}) reduces to the integral equation for the maximal backflow probability in free space~\cite{BM94Probability}, yielding $\mathcal{P}_{\max} = c_\text{bm}$.

\section{Summary and discussion}
\label{end}

In summary, we have considered classically-forbidden probability flow arising in two different physical scenarios: (i) the quantum backflow problem in which a wave packet comprised of non-negative momentum plane waves, Eq.~(\ref{bf-ini}), evolves under the action of a constant force, Eq.~(\ref{bf-SE}), and (ii) a DIT-type problem, called here the quantum reentry problem, in which a wave packet initially confined to the region of non-positive positions, Eq.~(\ref{dit-ini}), propagates in free space, Eq.~(\ref{dit-SE}). We have shown that the formula giving the backflow probability against a force $m g \ge 0$ during a time interval $(T_1,T_2)$, Eq.~(\ref{bf-P2}), is mathematically equivalent to the formula for the reentry probability at an observation point $\ell \ge 0$ during a time interval $(\tau_1, \tau_2)$, Eq.~(\ref{dit-P2}). Just like the maximal value of the backflow probability, $P_{\max}$, depends on a single dimensionless parameter $\alpha$, given by Eq.~(\ref{alpha}), the maximal value of the reentry probability, $\mathcal{P}_{\max}$, is determined by a single dimensionless parameter $\beta$, given by Eq.~(\ref{beta}). Based on the numerical investigation in Ref.~\cite{MB98velocity}, $P_{\max} \simeq c_\text{bm} e^{-2 \alpha}$ and $\mathcal{P}_{\max} \simeq c_\text{bm} e^{-2 \beta}$; for $\alpha = 0$ and $\beta = 0$, the values of both probabilities equal  $c_\text{bm}$.

The maximal backflow probability, $P_{\max}$, and the maximal reentry probability, $\mathcal{P}_{\max}$, decrease with increasing $\alpha$ and $\beta$, respectively. Equations~(\ref{alpha}) and (\ref{beta}) show that both $\alpha$ and $\beta$ increase with $m$ and decrease with $\hbar$, which means that both the backflow and reentry effects disappear in the (naive) classical limit of $m \rightarrow \infty$ or $\hbar \rightarrow 0$. Also, $\alpha$ increases (and $P_{\max}$ decreases) with $g$, and $\beta$ increases (and $\mathcal{P}_{\max}$ decreases) with $\ell$; this is in line with the intuitive expectation that the backflow and reentry effects must disappear in the limit of an infinitely large force and in the limit of an infinitely remote observation point, respectively.

In order to better understand the dependence of $P_{\max}$ and $\mathcal{P}_{\max}$ on the time intervals $(T_1,T_2)$ and $(\tau_1, \tau_2)$, respectively, we rewrite Eqs.~(\ref{alpha}) and (\ref{beta}) as follows:
\begin{equation*}
	\alpha = g \sqrt{\frac{m}{\hbar}} \, T \sqrt{\Delta T} \,,
\end{equation*}
where $T = \frac{1}{2} (T_1 + T_2)$ and $\Delta T = T_2 - T_1$, and
\begin{equation*}
	\beta = \ell \sqrt{\frac{m}{\hbar}} \sqrt{\frac{\Delta \tau}{\tau^2 - (\Delta \tau / 2)^2}} \,,
\end{equation*}
where $\tau = \frac{1}{2} (\tau_1 + \tau_2)$ and $\Delta \tau = \tau_2 - \tau_1$. These expressions reveal an interesting difference between the backflow and reentry effects. For $\Delta T$ fixed, the backflow effect becomes weaker as $T$ increases: $\alpha \sim T$, and so $P_{\max} \rightarrow 0$ as $T \rightarrow \infty$. On the contrary, the reentry effect becomes more pronounced with increasing $\tau$ (and $\Delta \tau$ fixed). Indeed, $\beta \sim 1 / \tau$ for large $\tau$, and so $\mathcal{P}_{\max} \rightarrow c_{\text{bm}}$ as $\tau \rightarrow \infty$.

While we have established mathematical equivalence between the integrated probability fluxes describing the quantum backflow and quantum reentry effects, the physics underlying this equivalence is still to be understood. For instance, it is important to construct compelling physical arguments explaining why the observation point $\ell$ in the reentry problem plays the role of the acceleration $g$ in the backflow problem, and why the maximal reentry probability $\mathcal{P}_{\max}$ increases with the increase of mean time $\frac{1}{2} (\tau_1 + \tau_2)$.

Finally, we believe that the equivalence between quantum backflow and quantum reentry, reported in this paper, has a potential to facilitate future experimental observations of a classically-forbidden probability flow: it might be easier to prepare an initial state with desired characteristics in the position space rather than in the momentum space.


%

\end{document}